\documentclass[preprint2]{aastex62}

\newcommand{\turbustat}{{\sc TurbuStat}}
\newcommand{\kms}{${\rm km\,s^{-1}}$}

\shorttitle{\turbustat: Turbulence Statistics in Python}
\shortauthors{Koch et al.}

\begin{document}


\title{\turbustat: Turbulence Statistics in Python}


\correspondingauthor{Eric W. Koch}
\email{koch.eric.w@gmail.com}
\author[0000-0001-9605-780X]{Eric W. Koch}
\author[0000-0002-5204-2259]{Erik W. Rosolowsky}
\affiliation{University of Alberta, Department of Physics 4-183 CCIS, Edmonton AB T6G 2E1, Canada}

\author[0000-0001-9857-1853]{Ryan D. Boyden}
\affiliation{Department of Astronomy and Steward Observatory, University of Arizona, 933 North Cherry Avenue, Tucson, AZ 85721, USA}

\author[0000-0001-5817-5944]{Blakesley Burkhart}
\affiliation{Center for Computational Astrophysics, Flatiron Institute, 162 Fifth Avenue, New York, NY 10010, USA}
\affiliation{Department of Physics and Astronomy, Rutgers, The State University of New Jersey, 136 Frelinghuysen Rd, Piscataway, NJ 08854, USA}

\author[0000-0001-6431-9633]{Adam Ginsburg}
\affiliation{Jansky fellow of the National Radio Astronomy Observatory, 1003 Lopezville Road, Socorro, NM 87801, USA}

\author{Jason L. Loeppky}
\affiliation{Department of Physics, University of British Columbia, Okanagan Campus, 3333 University Way, Kelowna, BC V1V 1V7, Canada}

\author[0000-0003-1252-9916]{Stella S.R. Offner}
\affil{Department of Astronomy, The University of Texas at Austin, 2515 Speedway, Stop C1400, Austin, TX 78712, USA}

\begin{abstract}
We present \turbustat\ (v1.0): a {\sc python} package for computing turbulence statistics in spectral-line data cubes. \turbustat\ includes implementations of fourteen methods for recovering turbulent properties from observational data. Additional features of the software include: distance metrics for comparing two data sets; a segmented linear model for fitting lines with a break-point; a two-dimensional elliptical power-law model; multi-core fast-fourier-transform support; a suite for producing simulated observations of fractional Brownian Motion fields, including two-dimensional images and optically-thin {\sc HI} data cubes; and functions for creating realistic world coordinate system information for synthetic observations.  This paper summarizes the \turbustat\ package and provides representative examples using several different methods. \turbustat\ is an open-source package and we welcome community feedback and contributions.
\end{abstract}

\keywords{turbulence --- methods: statistical --- methods: data analysis}

\section{Introduction} 
\label{sec:introduction}

Turbulence is ubiquitous throughout the interstellar medium \citep[see reviews by][]{Elmegreen2004ARA&A..42..211E,lazarian2009SSRv..143..357L}. Observations of different ISM phases demonstrate that similar turbulent properties are found over a wide range of scales \citep[e.g.,][]{Armstrong1995ApJ...443..209A,Chepurnov2010ApJ...710..853C}.  Connecting these observations with theoretical and numerical predictions is critical for understanding how turbulence affects the structure and motion of the ISM.  This connection provides important inputs for many astrophysical processes; for example, turbulence plays a central role in modern star formation theories \citep[e.g.,][]{Krumholz2009ApJ...699..850K,Ostriker2010ApJ...721..975O,Federrath2012ApJ...761..156F,Burkhart2018ApJ...863..118B}.

Determining turbulent properties from observations is challenging because of the limited information available and the complexity of the ISM.  Two-dimensional images, in particular column density or extinction maps, are a projection of a three-dimensional turbulent field.  Spectral-line observations provide additional constraints from the line-of-sight velocity, though features resolved in velocity may not be spatially-distinct \citep{Burkhart2013aApJ...771..122B,Beaumont2013ApJ...777..173B} and are affected by opacity and line excitation \citep{Burkhart2013ApJ...771..123B,Correia2016ApJ...818..118C}.  This means that observations miss up to four of the full six-dimensional phase-space\footnote{Three spatial and three velocity dimensions.} that describe the ISM structure and its motion.  The inherent complexity of the ISM adds to the difficulty in interpreting observations.  Turbulence in the ISM may be driven by multiple energy injection sources on different scales \citep{MO2007ARA&A..45..565M,2014prpl.conf..243K,Chepurnov2015ApJ...810...33C,krumholz2018} and is affected by variations in magnetic field strength and orientation \citep{Goldreich1995ApJ...438..763G,Cho2003,burkhart2009-bispec-moments,2014MNRAS.439.2197M,2015ApJ...805..118B,2017ApJ...842L...9H}. There are additional physical effects at work, including phase transitions that affect thermodynamic properties and gravitational collapse within molecular clouds \citep{bialy2017,2018ApJ...862...55H}.

Significant effort over the last $\sim40$ years has sought to connect predicted turbulent properties from theory \citep[e.g.,][]{Goldreich1995ApJ...438..763G} with (magneto-)hydrodynamic simulations \citep[e.g.,][]{Cho2003,2007ApJ...665..416K,Kowal2007,Federrath08,2012ApJ...750...13C} and observations \citep[e.g.,][]{Burkhart2010ApJ...708.1204B}.  To make these connections, an array of methods have been proposed in the literature to recover turbulent properties from observational data.  These methods utilize either two-dimensional images or three-dimensional spectral-line data cubes\footnote{Observations of polarization provide additional information \citep[e.g.][]{2011Natur.478..214G,2014A&A...566A...5I,2018ApJ...855...29H}, but \turbustat\ does not currently handle polarization data.}.  The latter has been of particular interest for recovering properties of the turbulent density and velocity fields \citep[e.g.,][]{lp00}.

We have developed \turbustat, a publicly-available Python package that implements fourteen observational diagnostics of ISM turbulence described in the literature. \turbustat\ provides a common framework for running and comparing turbulence diagnostics, including comparisons between simulations and observations \citep[e.g.,][]{Boyden2016ApJ...833..233B,Koch2017MNRAS.471.1506K,Boyden2018ApJ...860..157B,Haworth2018NewAR..82....1H}.  The use of some techniques has been limited by the lack of a publicly-available implementation.  Furthermore, many studies focus on using one or a small number of techniques.  This has resulted in a limited understanding of the regimes where particular methods are best-suited and the limits where inherent assumptions in a method break down.  The breadth of techniques in \turbustat\ provides the opportunity to explore these issues.

In this paper, we present an overview of \turbustat's first major release (v1.0), including a description of the methods implemented in \turbustat\ (\S\ref{sec:methods}), an overview of the package (\S\ref{sec:package_overview}), and a demonstration of \turbustat's capabilities with representative examples (\S\ref{sec:turbustat_examples}). The Appendices highlight the choice of normalization for the wavelet transform \citep[][Appendix \ref{app:issues_with_wavelet_normalization}]{gill-wavelet}, a comparison of our Delta-variance implementation to the original IDL code \citep[][Appendix \ref{app:comparison_to_deltavariance_idl_code}]{oss08I-delvar}, and a series of memory and timing tests for the methods in \turbustat\ (Appendix \ref{app:scaling_tests}).  \turbustat\ is open-source and includes extensive documentation and tutorials on the use of the methods (\url{turbustat.readthedocs.io}).  We encourage feedback from the community and welcome contributions.


\section{Methods}
\label{sec:methods}

\turbustat\ \citep[v1.0][]{turbustatv10} has implementations of 14 literature methods\footnote{We note that this number differs from \citet{Boyden2016ApJ...833..233B,Boyden2018ApJ...860..157B} since (i) the Tsallis statistic was not used and (ii) multiple outputs of dendrograms and statistical moments were counted as individual methods.} that recover properties related to turbulence from observational data.  We briefly describe the methods and relevant literature here \citep[also see descriptions in][]{Boyden2016ApJ...833..233B,Koch2017MNRAS.471.1506K,Boyden2018ApJ...860..157B}, and note that the package documentation contains thorough explanations and code examples.

\subsection{Structure Analysis}
\label{sub:struct_analysis}

The spatial structure of the ISM is hierarchical. Statistics that characterize the structural properties of an image or spectral-line data cube are one way to describe the hierarchical structure. \turbustat~has two methods which provide a non-parametric description of hierarchical structure.

\paragraph{\bf Genus} Genus statistics are a measure of topology. The value of the genus statistic is the difference between the number of isolated regions above and below a threshold.  A genus curve is produced by varying the threshold over a range of values.  The first use of the genus statistic on column density maps was introduced \citet{Lazarian2002ASPC..276..182L} and later expanded on by  \citet{Kowal2007}, \citet{Chepurnov2008} and \citet{Burkhart2012ApJ...749..145B}. The implementation in \turbustat\ closely follows the approach from \citet{Chepurnov2008}.

\paragraph{\bf Dendrograms} Dendrograms are a common method for exploring the hierarchical structure of data.  Their use in molecular cloud studies was proposed by \citet{dendrograms} and \citet{dendrograms-nature}, where pixels in an image or data cube are combined into hierarchical clusters based on their brightness \citep[also see][]{houlahan1,houlahan2}. \citet{burkhart-dendrograms} explored two statistics based on the dendrogram structure: (1) the relation between the number of structures in the dendrogram as a function of the branch height, and (2) the histogram of peak intensity in each structure of the dendrogram. Dendrograms can also be combined with other statistics, such as 1D PDFs \citep{2018ApJ...859..162C}. \turbustat\ implements both of these statistics and utilizes astrodendro\footnote{\url{dendrograms.readthedocs.io}} to compute the dendrograms.

\subsection{Properties of Turbulence}
\label{sub:prop_of_turb}

Despite the complexity of astrophysical turbulence, several statistics for observational data have theoretically-motivated properties. \turbustat\ implements a number of these methods, particularly those related to power-spectra.

\paragraph{\bf Spatial Power-spectrum (SPS)} The spatial power-spectrum is a widely-used method for finding the turbulent field index in the ISM.  The power-spectrum index from the column density or velocity centroid map is related to the underlying density or velocity field of the ISM, respectively \citep[e.g.,][]{lp00,Esquivel2005ApJ...631..320E}.  Most studies reduce the two-dimensional power-spectrum of the image into one dimension with azimuthal-averaging then fit a power-law to find the index \citep[e.g.,][]{Crovisier1983A&A...122..282C,Scalo1984ApJ...277..556S,stan01-sps,padoan-perseus,Burkhart2013ApJ...771..123B,2018ApJ...856..136P}. Additional information can be retained when modelling the full two-dimensional power-spectrum, including preferred directions of structure in the image and anisotropy \citep[e.g.,][]{Martin2015ApJ...809..153M,Kalberla2016A&A...595A..37K} that are predicted to relate to the magnetic field structure \cite[][]{Burkhart2014ApJ...790..130B,Kandel2017MNRAS.464.3617K,GonzalezCasanova2017ApJ...835...41G}. \turbustat\ can be used for both types of studies and can model breaks in the power-spectrum and the effect of a telescope beam (see \S\ref{sub:additional_features}).

\paragraph{\bf Modified Velocity Centroids (MVC)} MVC is an adaptation of the spatial power-spectrum introduced by \citet{Lazarian2003} that accounts for velocity-density correlations, which alter the power-spectrum of a velocity centroid map.  The centroid power-spectrum is corrected by subtracting the column density power-spectrum multiplied by the average velocity-dispersion in the data.  \citet{Esquivel2005ApJ...631..320E} explore the limits for when this correction is required.  The MVC implementation in \turbustat\ requires an input of the velocity centroid, integrated intensity\footnote{Or column density.}, and line width\footnote{From the second moment.} maps.  The formulation in \citet{Lazarian2003} requires {\it unnormalized} velocity centroids (i.e., without dividing by the integrated intensity). Since most observational products do not utilize this form of centroid, our implementation converts the centroid map to an unnormalized form. \turbustat's implementation of MVC has the same features described for the spatial power-spectrum.

\paragraph{\bf Velocity Channel Analysis (VCA)} \citet{lp00} show that, by considering the power spectrum of a spectral-line cube integrated over spectral frequencies, the index of the spatial power-spectrum will be altered by velocity fluctuations for sufficiently small spectral channel widths \citep[see also][]{lp04}.  By increasing the spectral channel width, the power-spectrum index approaches the power-spectrum of the column density, which is set only by density fluctuations \citep[e.g.,][]{stan01-sps,Muller2004ApJ...616..845M}.  Recent work by \citet{Kandel2016MNRAS.461.1227K} expands these theoretical predictions to include anisotropy in two-dimensional power-spectra.  The VCA implementation in \turbustat\ has the same functionality as the spatial power-spectrum and includes the ability to alter the spectral channel width of a given cube (see \S\ref{subsub:vca}).

\paragraph{\bf Velocity Coordinate Spectrum (VCS)}  The VCS is a complementary technique to VCA that is first mentioned in \citet{lp00} and later expanded on in \citet[][also see \citep{Lazarian2008ApJ...686..350L,chepurnov09}]{lp06}. While VCA integrates over the spectral dimension, VCS is the integration over the spatial dimensions, which yields a 1D spectral power spectrum.  The VCS implementation in \turbustat\ fits a broken linear model to the VCS (\S\ref{sub:additional_features}), mimicing the asymptotic high- and low-resolution solutions presented in \citet{lp06}.  A future extension of the code will include fitting with the complete VCS model \citep{chepurnov10,Chepurnov2015ApJ...810...33C}, which can, for example, constrain the turbulent driving scale.

\paragraph{\bf Bispectrum}  The bispectrum is the Fourier transform of the three-point correlation function, and is the next-order analog to the power-spectrum.  Unlike the power-spectrum, the bispectrum includes phase information, allowing for correlations between different spatial frequencies to be explored.  \citet{burkhart2009-bispec-moments} explored how these correlations change in different MHD regimes and later extended the study to consider the {\sc HI} column density of the SMC \citep{Burkhart2010ApJ...708.1204B}.  Quantitative comparisons between bispectra are more difficult than with the power-spectrum, which can be characterized only by its index.  The bispectrum is a complex quantity and cannot generally be reduced to a 1D representation.  Instead, the {\it bicoherence}---a real-valued normalized quantity that represents phase coupling---can be calculated.  The \turbustat\ implementation includes both quantities, using the bicoherence definition from \citet{Bicoherence-Hagihira}.  Calculating the bispectrum for even a small image is expensive and time-consuming.  Our implementation uses Monte Carlo sampling, with user input on the number of samples, to compute the bispectrum for each combination of wavenumbers.  A future extension to avoid this sampling is to utilize the multi-pole expansion introduced recently by \citet{Portilla2018ApJ...862..119P}.

\paragraph{\bf Wavelet Transform}  \citet{gill-wavelet} measure the amount of structure as a function of spatial scale by taking the sum of positive values in a wavelet decomposition of a two-dimensional image.  This gives a transform that is similar to the structure function \citep[e.g.,][]{miesch-bally}.  The wavelet implementation in \turbustat\ is similar to the algorithm from \citet{gill-wavelet}, though we introduce a change in the normalization of the wavelet kernels (Appendix \ref{app:issues_with_wavelet_normalization}).

\paragraph{\bf Delta-Variance}  Similar to the \citet{gill-wavelet} wavelet transform\footnote{Also see \citet{Zielinsky1999A&A...347..630Z}.}, the delta-variance technique is based on a wavelet decomposition, and is an extension of Allan Variance for one-dimensional time series \citep{stutzki98,bensch01-delvar,Ossenkopf2001A&A...379.1005O}.  The delta-variance characterizes the image structure at a set of spatial scales by calculating the variance in the wavelet decomposition.  An extension of this method for irregularly-shaped observational maps was developed by \citet{oss08I-delvar,oss08II-delvar}, which we have implemented in \turbustat\ (see Appendix \ref{app:comparison_to_deltavariance_idl_code} for a comparison with the IDL code provided by these authors).

\paragraph{\bf Principal Component Analysis (PCA)} PCA is a general dimensionality reduction procedure that identifies correlated components based on an orthogonal decomposition of a covariance matrix.  \citet{heyer-pca} first applied PCA to a spectral-line data cube by creating a covariance matrix of spectral channels in a data cube, decomposing that covariance matrix, and using the eigenvectors and eigenimages to recover characteristic spectral and spatial scales in the data.  Combining these scales over the first $N$ eigenvalues produces a size-line width relation for the data, whose index can be related to the theoretically expected turbulent regimes.  This technique was further developed in \citet{brunt-pca1,brunt-pca2} and \citet{RomanDuval2011ApJ...740..120R}, with an extension for measuring anisotropy \citep{anisotropy-pca}. An analytic model is presented in \citet{Brunt2013MNRAS.433..117B}. \turbustat\ implements the algorithm described in \citet{brunt-pca1,brunt-pca2}, including correction factors for the beam size and empirically-derived calibrations (see \S\ref{subsub:pca}).  The spectral and spatial scales can be fit with orthogonal distance regression or with a Bayesian approach, both of which handle errors in both dimensions.

\paragraph{\bf Spectral Correlation Function (SCF)} The SCF was introduced by \citet{scf} to relate spatial and spectral similarities of a data cube \citep[see also][]{padoan2001,padoan-scf}.  \turbustat\ implements the form from \citet{Yer14}, where the statistic is the normalized root-mean-square difference between the cube and its spatially-shifted self.  By iterating over a range of spatial shifts in both spectral dimensions, we create a two-dimensional correlation surface whose azimuthally-averaged index has been shown to be sensitive to changes in turbulent properties \citep{padoan-scf,Muller2004ApJ...616..845M,Gaches2015ApJ...799..235G}.  Our implementation can model the correlation surface in either one- or two-dimensions using a similar approach described for the spatial power-spectrum.

\subsection{Analysis of Distributions}
\label{sub:distrib_analysis}

The distribution of values within a data set, or portions of a data set, are useful diagnostics in many settings. \turbustat\ provides a convenient implementation for fitting probability distribution functions, and descriptions of distribution shapes.

\paragraph{\bf Probability Distribution Functions (PDF)}  The most commonly-used analysis technique to describe turbulent properties from observational data products is the PDF.  Extensive work on PDFs from simulations \citep[e.g.,]{1994ApJ...423..681V,Ostriker2001ApJ...546..980O,Kowal2007,Federrath08,burkhart2009-bispec-moments,Burkhart2017ApJ...834L...1B} and observations \citep[e.g.,][]{Miesch1995ApJ...450L..27M,Burkhart2010ApJ...708.1204B,Lombardi2015,Imara2016ApJ...829..102I,bialy2017} has provided a solid framework connecting the PDF to turbulent properties. The \turbustat\ PDF implementation was written to emphasize flexibility in treatment and modelling of PDFs.  Both images and cubes can be used, and the code can be used to quickly recover properties of the PDF and its empirical cumulative distribution function (ECDF). The implementation utilizes the {\sc scipy.stats}\footnote{\url{docs.scipy.org/doc/scipy/reference/stats.html}} continuous distributions for modelling, including normal, log-normal, and power-law distributions. A maximum-likelihood estimator is used to fit model distributions to the data.

\paragraph{\bf Statistical Moments} An extension to PDF studies is an analysis of higher-order moments, namely the skewness and kurtosis, which are non-Gaussian indicators. Previous studies have explored skewness and kurtosis for column density, velocity centroid, and line width PDFs of simulations and observations \citep[e.g.,][]{Padoan1999ApJ...525..318P,Kowal2007,burkhart2009-bispec-moments,Burkhart2013ApJ...771..123B,Burkhart2015ApJ...808...48B}. \citet{Burkhart2010ApJ...708.1204B} extend this approach by using a rolling circular filter to create spatial moment maps of the {\sc HI} in the SMC.  This allows for spatial variations in the PDF moments to be explored.  The \turbustat\ implementation provides both methods.

\paragraph{\bf Tsallis Statistics} The Tsallis distribution was introduced by \citet{tsallis1988} for describing multi-fractal systems.  \citet{Esquivel2010ApJ...710..125E} first introduced Tsallis statistics for modelling ISM turbulence \citep[see also][]{Tofflemire2011,Burkhart2013aApJ...771..122B,Burkhart2015ApJ...808...48B,GonzalezCasanova2018MNRAS.475.3324G}. Our implementation of Tsallis statistics follows \citet{Esquivel2010ApJ...710..125E} and uses a q-Gaussian distribution to model the difference in a given image as a function of spatial scale.

\section{Overview of \turbustat} 
\label{sec:package_overview}

\turbustat\ contains sub-modules for the computation of the methods, calculating moment arrays and their uncertainties from data cubes, and basic I/O operations for handling FITS files.  In this section, we present the methods and other helpful utilities implemented in \turbustat.

\subsection{Package Dependencies}

\turbustat\ utilizes several packages in the python scientific computing infrastructure, namely built on numpy \citep[][\url{numpy.org}]{numpy}, scipy \citep[][\url{scipy.org}]{scipy}, and matplotlib \citep[][\url{matplotlib.org}]{mpl}. We use astropy \citep[][\url{astropy.org}]{astropy2} for I/O operations, unit handling, convolution, and WCS transformations. The package infrastucture of \turbustat\ relies heavily on the astropy testing and documentation infrastucture\footnote{\url{github.com/astropy/astropy-helpers}}. The scikit-image \citep[][\url{scikit-image.org}]{scikit-image} is used for morphological operations, contour finding, and fitting elliptical models. The scikit-learn \citep[][\url{scikit-learn.org}]{scikit-learn} provides a parallelized routine for calculating pairs of distances between data sets, which is required for the Cramer distance metric \citep[][\S\ref{sub:distance_metrics}]{Yer14,Koch2017MNRAS.471.1506K}. Most of the fitting routines in \turbustat\ rely on the statsmodels package \citep[][\url{statsmodels.org}]{statsmodels}, which mimics the linear model fitting in the R programming language\footnote{\url{r-project.org}}. We also use statsmodels for maximum likelihood estimation for fitting PDFs. The astrodendro\footnote{\url{dendrograms.readthedocs.io}} package creates a dendrogram to explore hierarchical structure \citep{dendrograms-nature,dendrograms}; \turbustat\ creates statistical descriptions of the dendrogram.

\turbustat\ also has a few optional dependencies. The spectral-cube \citep[][\url{spectral-cube.readthedocs.io}]{sccube} and radio-beam\footnote{\url{radio-beam.readthedocs.io}} packages provide convenient methods for handling large data sets and beam manipulation, respectively. The emcee \citep[][\url{dfm.io/emcee}]{emcee} package provides optional MCMC fitting for the size--line width relation in PCA and distribution fitting for PDFs. The corner package \citep[][\url{github.com/dfm/corner.py}]{corner} creates convenient summary plots of the emcee sampler. For taking the Fast Fourier Transform (FFT) of large data, the FFTW library \citep[][\url{fftw.org}]{FFTW05} through the pyFFTW wrapper \citep[][\url{hgomersall.github.io/pyFFTW}]{pyfftw} can calculate the FFT in parallel.

\subsection{Methods Implementation}
\label{sub:methods_implementation}

The 14 turbulence methods implemented in \turbustat\ are implemented using a common framework to facilitate ease-of-use (the methods are presented in \S\ref{sec:methods}).  Each method is implemented as a python class with a common set of steps:

\begin{enumerate}
    \item {\bf Input} -- The data, FITS header and other relevant input information (i.e. beam size, distance to region) are given as inputs when initializing the method class.
    \item {\bf Calculation} -- Functions are defined in the class that performs the analysis.  Depending on the complexity of the method, the computing steps are split into multiple parts.  For example, the spatial power-spectrum has separate steps to (i) compute the power-spectrum, (ii) fit the 1D power-spectrum, (iii) fit the 2D power-spectrum, and (iv) produce a summary plot and print fit statistics.
    Running each of these functions in the order given will compute the entire method.  This step-by-step method allows for maximum user input when computing the method as the arguments and keyword arguments of each function can be altered.
    \item {\bf All-in-one} --  The multi-step approach described above can be cumbersome and requires remembering each step for computing the method.  For ease-of-use, each method includes a \texttt{run} function that runs all of the steps with sensible default settings and optionally returns a figure summarizing the results.  This approach is ideal for quickly computing a method for exploratory analyses.  Most key settings can be altered in the \texttt{run} function so that users need only use this function for most cases.
    \item {\bf Plotting \& Summary} -- Each method has a separate function that returns a summary plot of the method, including fits to the outputs of the method.
\end{enumerate}

Some methods have additional functions defined that return useful information about the method or data.  For example, after computing the PDF of a data set, the PDF class has functions for returning the percentile of a given value in the data.

\subsection{Distance Metrics} 
\label{sub:distance_metrics}

\turbustat\ includes distance metrics that use an output of a method to compare two data-sets. The methods implemented in \turbustat\ measure properties of the underlying physics in observational data; these distance metrics are one approach for quantifying the difference in their physical properties. In previous works, we have used these distance metrics to find which methods are sensitive to different input parameters in sets of simulations \citep{Yer14,Boyden2016ApJ...833..233B,Koch2017MNRAS.471.1506K,Boyden2018ApJ...860..157B}.  We refer readers to \citet{Koch2017MNRAS.471.1506K} for a full description of the distance metrics.  The \turbustat\ documentation also includes tutorials on the use of the distance metrics.


\subsection{Data Structure and Utilities} 
\label{sub:data_structure}

\paragraph{{\bf Data Structure}} \turbustat\ is primarily intended to work with observational data products, namely two-dimensional images and spectral-line data cubes.  However, our goal is also to ensure these methods are easily-used with a wide-array of data.  As such, the \turbustat\ methods accept two main input types: (1) Utilizing the astropy I/O interface, the methods in \turbustat\ expect a FITS HDU as input. (2) Data can also be passed as a numpy array when converting to the FITS format is prohibitive.  These inputs still require a FITS header be provided (see below).  Data types from the spectral-cube package may also be used.  The FITS header is required by most \turbustat\ methods for converting pixel scales into sky coordinates or the spectral dimension in a data cube.

\paragraph{\bf Generating FITS headers} Non-observational data, such as simulated observations, may not be saved with mock WCS information included.  For these cases, \turbustat\ includes utility functions to generate a FITS header or HDU for the data.

\paragraph{\bf Spectral Moments} Some methods in \turbustat\ require a data cube while others need a two-dimensional image representing some property of the data (i.e. integrated intensity or column density).  \turbustat\ includes utilities for calculating two-dimensional moment images from a data cube---namely the zeroth (integrated intensity), first (centroid) and second (line width) moments---using the spectral-cube package, and saving the images as FITS files.  Uncertainty maps for the moments can also be calculated, which are useful for down-weighting noisy regions in some methods.


\subsection{Generating fBM images and data cubes}
\label{sub:generating_fbm_images_and_data_cubes}

Fractional Brownian Motion (fBM) fields are useful, highly-simplified versions of a turbulent field. They are created by setting the power-law amplitude and randomly-drawing phases in the Fourier domain before transforming to real space.  \turbustat\ includes routines for creating 2D or 3D fBM fields and routines for creating mock optically-thin {\sc HI} data cubes given a 3D density and velocity field.

Two-dimensional fBM images can be created with a given power-law index, ellipticity (anisotropy), and angle of ellipticity.  These routines are primarily used in \turbustat\ to test methods but also provide useful examples (\S\ref{sub:examples_with_twodim_images}).

Three-dimensional fBM fields can also be generated, though the current implementation does not include the option for creating anisotropic fields.

Utilizing the aforementioned three-dimensional fields, \turbustat\ can generate mock optically-thin {\sc HI} data cubes from a given set of density and velocity fields.  These mock data cubes are another useful tool for testing, and an example is shown in \S\ref{sub:example_with_cubes}.

\subsection{Additional Features}
\label{sub:additional_features}

\turbustat\ has a number of additional features that we briefly list here. We note that the use of these features is well-described in the documentation with complete examples.

\begin{enumerate}
    \item {\it Beam corrections} --- Most astronomical imaging, particularly in the radio and submillimetre, are over-sampled relative to the area of a resolution element, or the beam size.  This leads to systematic correlations on scales of order the beam size or smaller that affects most turbulence methods that utilize spatial information.  For example, the spatial power-spectra will steepen on scales similar to the beam size, approaching the power-spectrum of the beam.  \turbustat\ includes routines for correcting the beam response in power-spectrum methods (SPS, MVC, VCA) that rely on the radio-beam package.  The PCA implementation includes the beam correction described in \citet{brunt-pca1}. For other spatial methods where the transform remains in real-space (wavelet, delta-variance, statistical moments, SCF), we recommend closely examining the response on scales small or near the beam size, and possibly avoiding those region when fitting.
    \item {\it Apodizing kernels} --- The power-spectrum methods (SPS, MVC, VCA) use an FFT that will exhibit the Gibbs phenomenon (ringing) when there is signal at the edge of the image.  The edges of an image can be tapered to avoid this ringing by using an apodizing kernel. \turbustat\ includes several options, with a tutorial demonstrating their effect on the power-spectrum, based on the routines implemented in photutils \citep[][\url{photutils.readthedocs.io}]{photutils}.
    \item {\it Parallelized FFTs} --- Computing the fast-fourier transform (FFT) is a bottleneck when running FFT-based methods (SPS, MVC, VCA, wavelets, Delta-variance) on large data sets.  To speed the FFT up, pyFFTW \citep[][\url{hgomersall.github.io/pyFFTW}]{pyfftw} can optionally be used to run the FFT in parallel.
    \item {\it Segmented Linear Model} --- \turbustat\ implements a segmented linear model described in \citet{muggeo2003estimating} that fits for a piece-wise linear model and the position of the breaks.  This model is used by default for the VCS, and can be optionally used for all power-spectrum-based methods.  For example, this model can be used to constrain break-points in spatial power-spectra that are related to a galaxy's disk scale height \citep[e.g.,][]{Combes2012A&A...539A..67C}.
    \item {\it 2D Elliptical Power-law Model} --- \turbustat\ includes an elliptical power-law model, adapted from \citet[][Eqs.  $3\mbox{--}5$]{Tessore2015A&A...580A..79T}, for fitting 2D power-spectra and constraining anisotropy.  The model is defined by three parameters: the power-law index, the ellipticity, and the elliptical angle.  Ellipticity parameterizes anisotropy and is defined such that $0$ is infinitely anisotropic and $1$ is isotropic.  The elliptical angle is the angle between the $x$-axis and the direction of the anisotropy.  An example of this model is shown in \S\ref{subsub:anisotropy_in_fourier_methods}.  The \turbustat\ implementation fits the logarithm of the 2D power-spectrum and estimates uncertainties using residual bootstrapping.
\end{enumerate}


\section{\turbustat \, Examples} 
\label{sec:turbustat_examples}

This section presents examples of the methods in \turbustat\ with idealized synthetic observations (\S\ref{sub:generating_fbm_images_and_data_cubes}).  We highlight the code's ability to recover expected parameters for methods that utilize data cubes (\S\ref{sub:example_with_cubes}) and two-dimensional images (\S\ref{sub:examples_with_twodim_images}). Scripts to reproduce these examples are available at \url{https://github.com/Astroua/TurbuStat/tree/master/Examples}.

\subsection{Examples with Position-Position-Velocity Cubes} 
\label{sub:example_with_cubes}

\subsubsection{Example Cubes}
\label{subsub:example_cubes}

For the following examples, we generate four idealized spectral-line data cubes.  These cubes are generated from three-dimensional fBM fields with a shape of $256^3$ and an index of $-4$ for the density and velocity fields.  We assume an isotropic velocity and so only generate one component of the velocity.  The velocity field has a dispersion of $10$ \kms, and the density field has a dispersion of $1$ cm$^{-3}$.  The set of four fields differ only in the random seed used to generate their phases.

The density fBM field is not positive definite and must be altered to ensure that it is.  Many approaches have been proposed to create a positive definite field from a fBM field, including taking the absolute value \citep{stutzki98}, taking the exponential to generate a log-normal distribution \citep{brunt-pca1,Ossenkopf2006A&A...452..223O,RomanDuval2011ApJ...740..120R}, subtracting the minimum value of the field \citep{MivilleDeschenesD2003ApJ...593..831M} or some multiple of the field's standard deviation \citep{Ossenkopf2006A&A...452..223O}.  We mimic the latter approach by adding the standard deviation of the density field to itself and setting values that remain negative to zero.  This distorts the index of the density field's power-spectrum index, however, the measured indices vary by less than $0.1$ from the original index of $-4$.

\begin{figure*}
\includegraphics[width=0.9\textwidth]{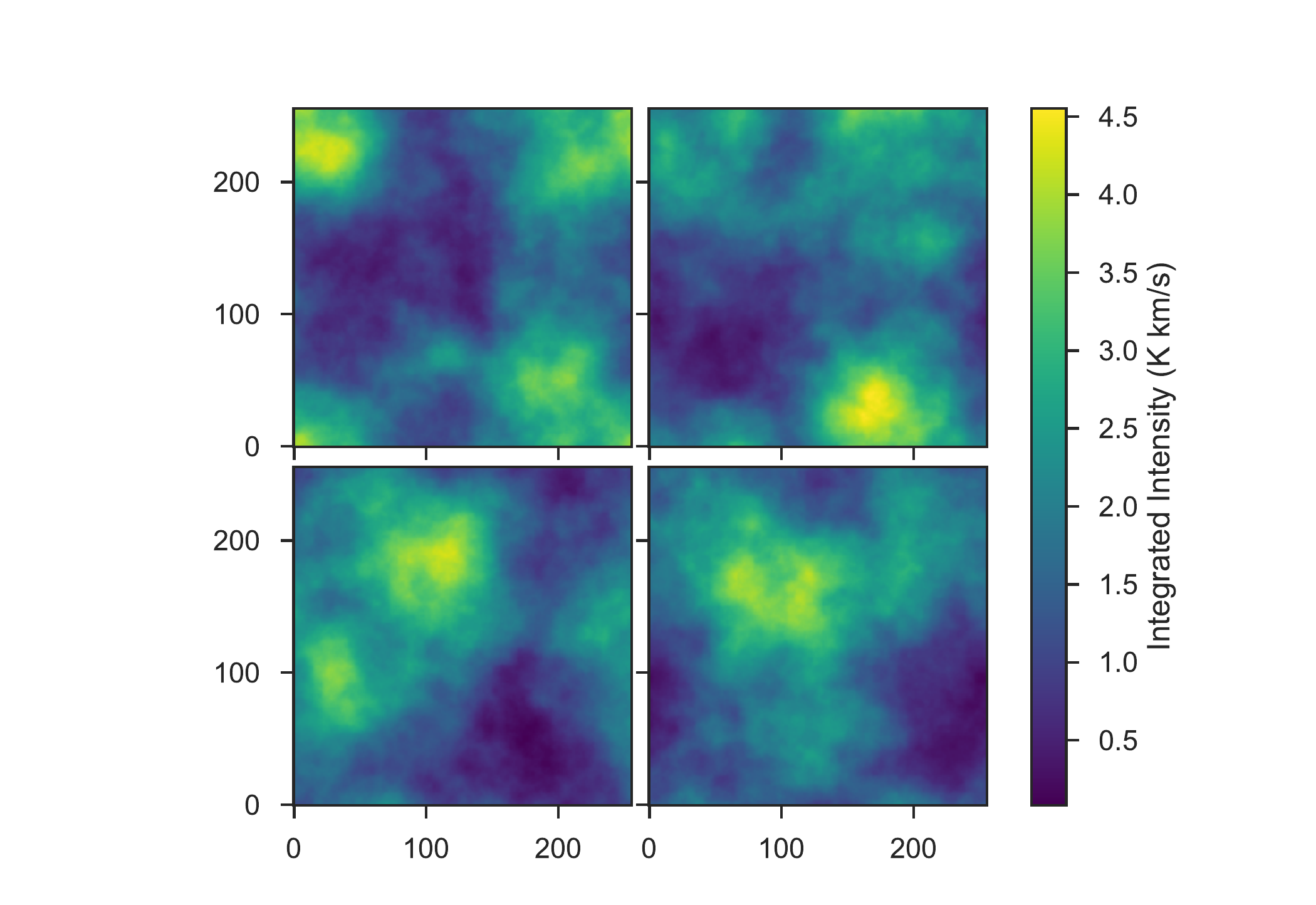}
\caption{\label{fig:cube_example} Integrated intensity maps of the idealized fBM {\sc HI} cubes used in \S\ref{sub:example_with_cubes}. We generated the cubes from density and velocity fields with indices of $-4$ that differ only in their random phases.}
\end{figure*}

We then generate four spectral-line data cubes of {\sc HI} emission assuming optically-thin conditions and a temperature of 100~K (the thermal velocity dispersion $\sigma_{\rm therm}=0.76$~\kms).  We set the velocity channel width to be $0.2$~\kms, and thus expect the spectral line profiles to be smooth.   Adopting a thermal line width is necessary for smoothing away ``shot noise'' on small scales \citep{Lazarian2001ApJ...555..130L}.  Figure \ref{fig:cube_example} shows the integrated intensity images of the four cubes.

We note that these are small cubes that have been generated for these examples.  We have chosen to do this to make reproducing these examples computationally inexpensive.  However, the choice to use small cubes does impose limitations.  The aforementioned ``shot noise'' that arises from having a finite number of emitters along the line-of-sight is a significant problem here \citep{Lazarian2001ApJ...555..130L}. \citet{Esquivel2003MNRAS.342..325E} and \citet{chepurnov09} demonstrate that this numerical effect drastically affects the recovered indices for the VCA and VCS.  By selecting a steep field index ($-4$) and by smoothing the data cubes with the thermal line width ($\sigma_{\rm therm}=0.76$~\kms), we ensure that velocity slices remain close to the thin VCA regime on scales larger than $2$~pixels \citep[based on Eqns. 12--14 from][]{Esquivel2003MNRAS.342..325E}.

We also note that these idealized data cubes have no density-velocity correlations \citep{Esquivel2007MNRAS.381.1733E} and are not affected by absorption or optical-depth effects \citep{lp04,Burkhart2013ApJ...771..123B}.

\subsubsection{Velocity Channel Analysis}
\label{subsub:vca}

\citet{lp00} predict that the spatial power-spectrum index will change as a function of the velocity channel width due to differences in the influence of the underlying velocity and density fields.  As the channel width increases, velocity fluctuations are averaged out and the power-spectrum index approaches the index of the density field.  We have chosen the VCA as an example because the analytic relation for the index with spectral channel width provides a good test case to ensure the \turbustat\ implementation, and the mock data cube creation, is correct.

\begin{figure*}
\plotone{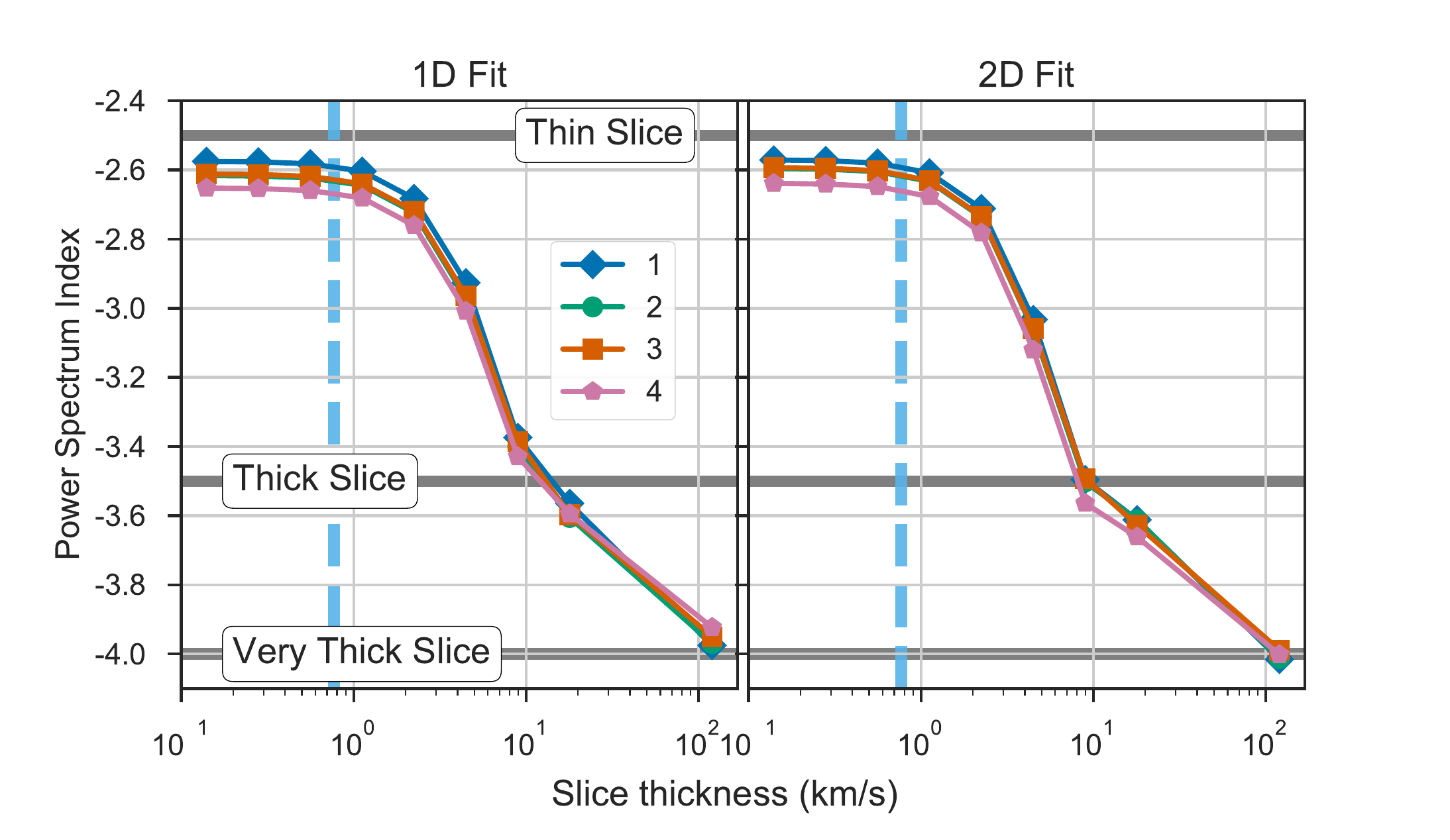}
\caption{\label{fig:vca_figure} Measured VCA index from fits to the 1D (left) and 2D power-spectra (right) as a function of velocity slice thickness for four example cubes generated with different random seeds (\S\ref{subsub:example_cubes}).  The vertical line is the thermal velocity dispersion used for smoothing, and the horizontal lines are the predicted slopes in the thin ($-2.5$), thick ($-3.5$) and very thick velocity ($-4.0$) regimes for density and velocity fields with an index of $-4$.  Shot noise and smoothing with the thermal velocity dispersion result in not recovering the thin velocity regime \citep{Esquivel2003MNRAS.342..325E,chepurnov09}. Indices fitted to the 2D power-spectrum are moderately steeper than the 1D power-spectrum indices due to the larger number of samples on small scales (large frequencies) in the 2D power-spectrum. The \turbustat\ implementation of the VCA and spatial power-spectrum recover the expected indices.}
\end{figure*}

There are three regimes for the VCA index described by \citet{lp00}, (1) the ``thin'' velocity regime dominated by velocity fluctuations: $m=-3 - (\gamma_v + 3)/2$, where $m$ is the power-spectrum index and $\gamma_v$ is the velocity field index; (2) the ``thick'' velocity regime where most velocity fluctuations have been averaged over: $m=-3 + (\gamma_v + 3)/2$; and (3) the ``very thick'' velocity regime where all velocity channels have been integrated over: $m=\gamma_n$, where $\gamma_n$ is the density field index.  Note that these relations are for the {\it steep} density regime ($\gamma_n < -3$); see \citet{lp00} for the {\it shallow} density regime.  For the $\gamma_v=\gamma_n=-4$ for these examples, we expect $m_{\rm thin}=-2.5$, $m_{\rm thick}=-3.5$, and $m_{\rm very thick}=-4$, which are shown as horizontal lines in Figure \ref{fig:vca_figure}.

Using \turbustat's VCA and spatial power-spectrum implementations, we fit one- and two-dimensional power-law models to the power-spectra as a function of channel width.  Figure \ref{fig:vca_figure} shows the recovered indices of the four example cubes.  For thin velocity channels, the recovered indices are $\sim\ -2.6$ and do not approach the expected $-2.5$ in the thin regime due to shot noise and smoothing from the thermal line width, shown with the blue vertical line.  For larger channel widths, we recover the expected indices in the thick and very thick regimes.  The two-dimensional power-law indices are moderately steeper in the thick velocity regime as these fits are dominated by the larger number of samples at large spatial frequencies. The variation in indices between the four cubes is small ($<0.1$).

\subsubsection{Principal Component Analysis}
\label{subsub:pca}

\begin{figure}
\includegraphics[width=0.5\textwidth]{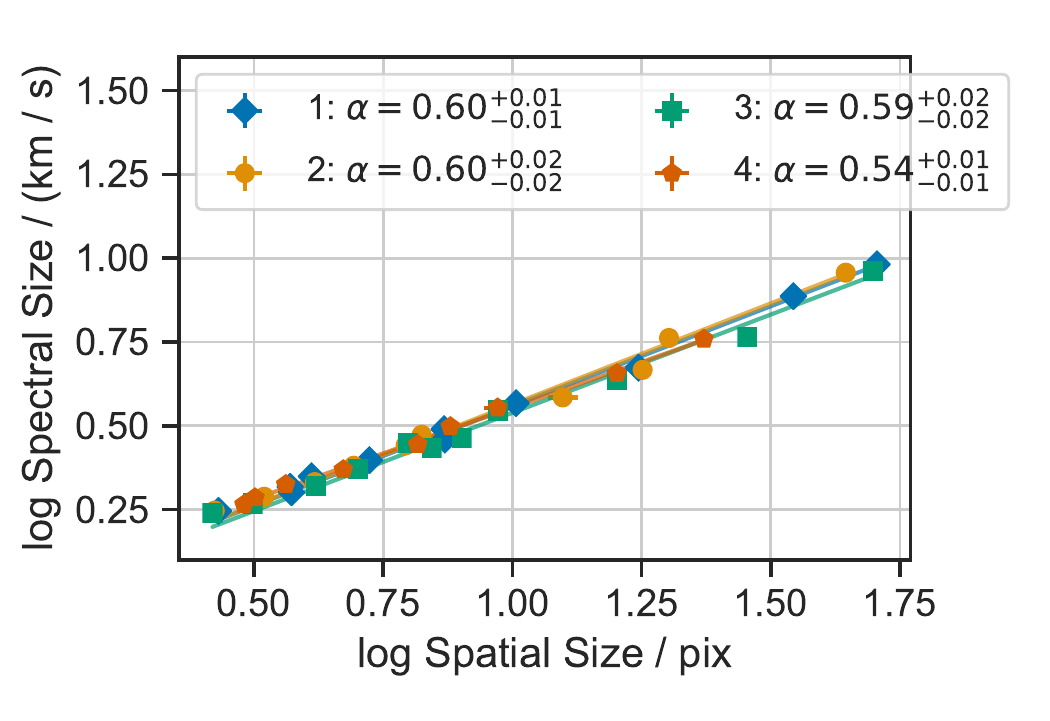}
\caption{\label{fig:pca_figure} Size-line width relations from PCA for the four example cubes.  The expected index of the size-line width relation for a velocity field with index $-4$ is $1/2$; the fitted indices are consistently steeper than the expected value.  These results demonstrate that similar slopes are consistently recovered with the \turbustat\ PCA implementation.}
\end{figure}

We also demonstrate our implementation of PCA on the four example cubes. Adopting a minimum eigenvalue of $0.001$ to avoid numerical noise in the PCA decomposition, we calculate the spatial and spectral scales and fit the size-line width relation for the four cubes, shown in Figure \ref{fig:pca_figure}.  For this example, we fit the size-line width relations with orthogonal distance regression\footnote{Implemented in scipy.}.

For a velocity field index of $-4$, we expect the size-line width relation to have an index of $1/2$.  The recovered indices are consistently higher than $1/2$, which we expect is due to to a small number of eigenvalues containing useful information, and three of the cubes have indices consistent within the uncertainty.  The fourth cube has a smaller index---closer to the expected index---than the other cubes which may be due to the autocorrelation surface of the first eigenimage not containing the $1/e$ contour used to find the spatial scale.  This issue has been noted in other PCA works \citep[e.g.,][]{RomanDuval2011ApJ...740..120R}. We also note that no correction factor has been used here, as is typically done to relate the measured index to the index of the energy spectrum \citep{brunt-pca1,RomanDuval2011ApJ...740..120R}.  The aforementioned works impose a log-normal distribution on the density fBM field, which is not done here, and the applicability of these correction factors may not be suitable for this case.  However, \turbustat\ includes the option to apply the correction factor from \citet{brunt-pca1}.

\subsection{Examples with two-dimensional images}
\label{sub:examples_with_twodim_images}

We demonstrate some of the methods which apply to two-dimensional images in this section, namely the recovery of known indices from fBM images and fitting elliptical power-law models to constrain anisotropy.

\subsubsection{Image Index Recovery}
\label{subsub:image_index_recovery}

We generate a series of two-dimensional fBM images with indices ranging from $\beta=0.5\mbox{--}4$ and different random seeds to demonstrate how well the delta-variance, spatial power-spectrum, and wavelets recover the indices.  Each of these methods provides complementary information based on their fitted slopes.  The delta-variance and wavelet transforms are similar to a second-order structure function and their measured slopes are related to the power-spectrum index with $\beta = m_{\rm delta-variance} + 2$ and $\beta = \left( m_{\rm wavelet} + 1 \right) / 2$, respectively.

\begin{figure}
\includegraphics[width=0.48\textwidth]{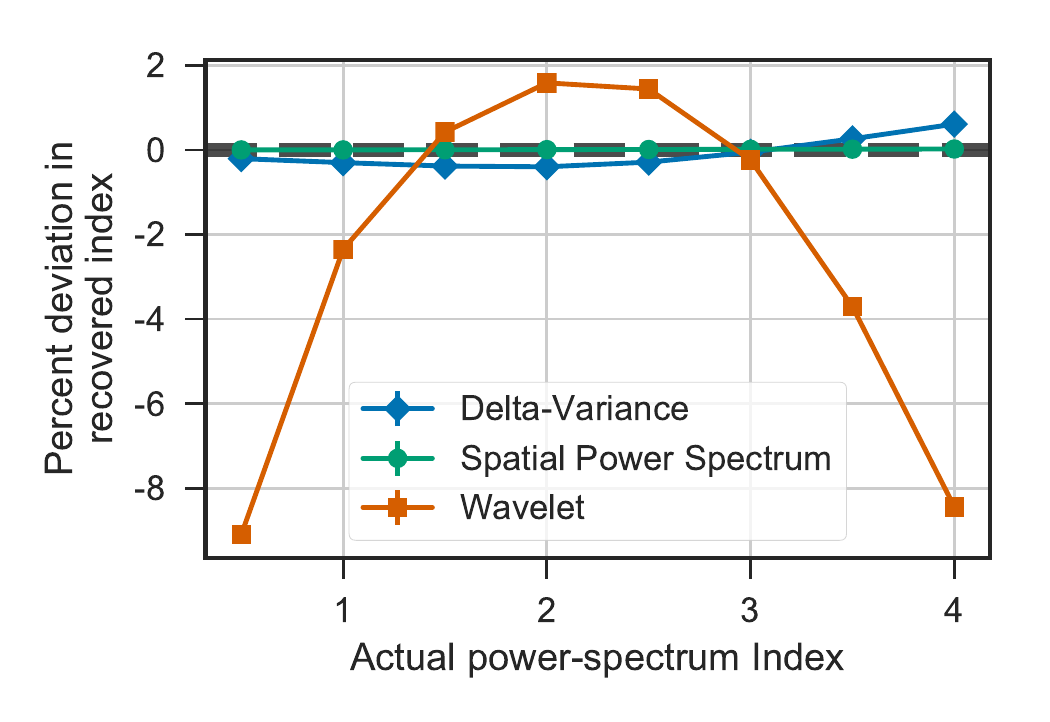}
\caption{\label{fig:2D_index_recovery} Percent deviation between the measured and actual power-spectrum index for fBM images with the delta-variance (blue diamonds), spatial power-spectrum (green circles), and wavelet (orange squares) methods. The former two methods accurately recover the power-spectrum index to within a few percent, which for observational data will likely be smaller than the uncertainty.  The wavelet method shows larger deviations that depend strongly on the index of the image.  Over the range of indices expected for interstellar turbulence, the delta-variance and spatial power-spectrum accurately recover the index of an image.}
\end{figure}

Figure \ref{fig:2D_index_recovery} shows the percent deviation of the measured index from the actual index across the range of fBM images. The spatial power-spectrum recovers the correct index in each case, while the index from the delta-variance deviates by $<1\%$.  The index from the wavelet transform, however, has larger systematic deviations that vary across the range of fBM image indices shown.  These deviations suggest that the delta-variance and spatial power-spectrum provide more accurate results than the wavelet transform.

\subsubsection{Modelling spatial anisotropy}
\label{subsub:anisotropy_in_fourier_methods}

There is significant interest in studying the connection between intensity and velocity anisotropy with magnetic field structure in the ISM \citep[e.g.,][]{Esquivel2005ApJ...631..320E,Burkhart2014ApJ...790..130B,Kalberla2016A&A...595A..37K,Kandel2017MNRAS.464.3617K}.  In this example, we demonstrate \turbustat's implementation of a two-dimensional elliptical power-law model for fitting power-spectra and other two-dimensional surfaces.  The model is described in \S\ref{sub:additional_features}.

\begin{figure*}
\includegraphics[width=0.9\textwidth]{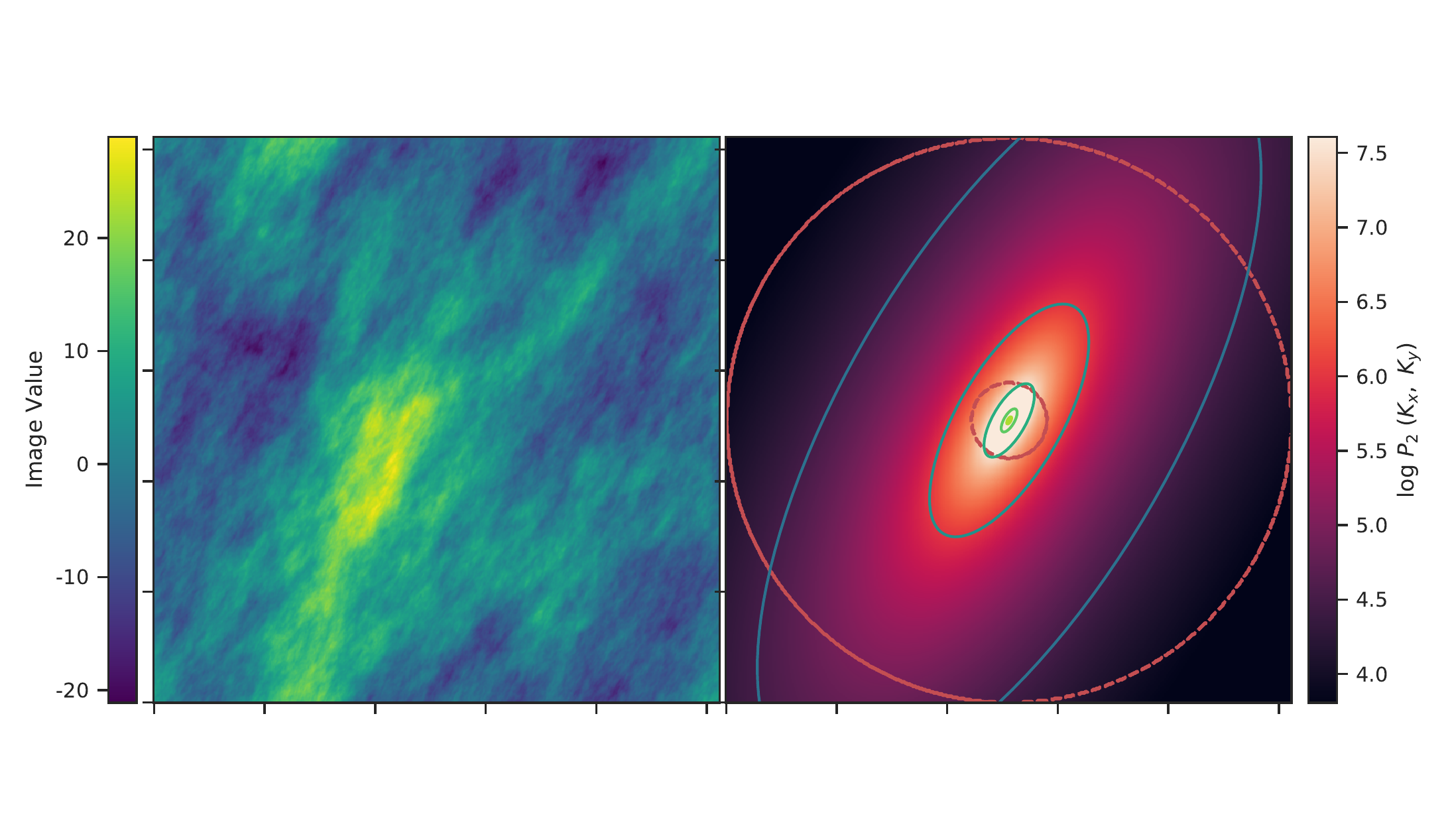}
\caption{\label{fig:2Dindex_anisotropy} Left: Anisotropic fBM image with index $-3$, ellipticity $0.4$, and angle $60^\circ$ (with respect to the $x$-axis).  Right: Two-dimensional power-spectrum (image) with an elliptical power-law model (solid contours).  The region between the dashed red contours is the data used in the fit. The model recovers all three parameters of the anisotropic fBM image to within the fit uncertainties.}
\end{figure*}

The left panel of Figure \ref{fig:2Dindex_anisotropy} shows a fBM field with a slope of $-3$ and ellipticity of $0.4$\footnote{Defined such that an ellipticity of $1.0$ is isotropic and $0.0$ is infinitely anisotropic.} oriented $60\degr$ above the $x$-axis in the figure.  Using the spatial power-spectrum, we find the two-dimensional power-spectrum shown in the right panel, which highlights the anisotropy of the structure.  Contours of the fitted elliptical power-law are shown with solid lines in the figure.  The model correctly recovers the aforementioned fBM image parameters.

This model can be used with other power-spectrum-based methods in \turbustat, including MVC and VCA, and the SCF correlation surface to constrain anisotropy in images.


\section{Conclusions} 
\label{sec:conclusions}

We introduce \turbustat, a Python package for turbulence statistics.  Currently, the package includes 14 methods presented in the literature for recovering turbulent properties from ISM observations.  This paper describes the capabilities of the implementations and other utilities in the package.  We also present a few examples (\S\ref{sec:turbustat_examples}) that demonstrate the code's ability to recover expected properties.  We note that additional turbulence data sets in FITS format can be obtained online at \url{www.mhdturbulence.com} as part of the Catalog for Astrophysical Turbulence Simulations (CATS) project.  CATS also includes spectral line data compatible with \turbustat.

While this version of \turbustat\ includes many methods for recovering turbulent properties, there are a number of other methods that we hope to include in the future. Some examples include: structure functions \citep[e.g.,][]{Boldyrev2002ApJ...573..678B,Padoan2003ApJ...583..308P}, the phase coherence index \citep{Burkhart2016ApJ...827...26B}, the brightness distribution index \citep{Sawada2012ApJ...752..118S}, the velocity gradient technique  \citep[e.g.,][]{Yuen2017ApJ...837L..24Y,Lazarian2018ApJ...865...46L}, wavelet-based cross-correlation analysis \citep{Arshakian2016A&A...585A..98A}, and complete VCS modelling \citep{chepurnov10,Chepurnov2015ApJ...810...33C}.

\turbustat\ is open-source\footnote{\url{github.com/Astroua/TurbuStat}} and has thorough documentation and tutorials available\footnote{\url{turbustat.readthedocs.io}}. We welcome feedback, recommendations, and contributions from the community to improve the on-going development of \turbustat.


\acknowledgments

We thank Caleb Ward for important contributions to the code in its early development.  We are grateful for feedback and issues reported by Dario Colombo, Jesse Feddersen, Simon Glover, Jonathan Henshaw, Andr\'{e}s Izquierdo, and Sac Medina.  We thank the anonymous referee for their careful reading of the manuscript and comments.  EWK is supported by a Postgraduate Scholarship from the Natural Sciences and Engineering Research Council of Canada (NSERC). EWK and EWR are supported by a Discovery Grant from NSERC (RGPIN-2012-355247; RGPIN-2017-03987).  This research was enabled in part by support provided by WestGrid (\url{www.westgrid.ca}), Compute Canada (\url{www.computecanada.ca}), and CANFAR (\url{www.canfar.net}).

\software{astropy \citep{astropy2}, matplotlib \citep{mpl}, seaborn \citep{seaborn}, numpy \citep{numpy}, scipy \citep{scipy}, scikit-image \citep{scikit-image}, scikit-learn \citep{scikit-learn}, statsmodels \citep{statsmodels}, astrodendro (\url{dendrograms.readthedocs.io}), spectral-cube (\url{spectral-cube.readthedocs.io}), radio-beam (\url{radio-beam.readthedocs.io}), emcee \citep{emcee}, corner \citep{corner}, pyFFTW \citep{pyfftw} and FFTW \citep{FFTW05}}

\bibliographystyle{aasjournal}
\bibliography{astrostat}

\appendix

\section{Issues with Wavelet Normalization} 
\label{app:issues_with_wavelet_normalization}

When implementing the wavelet transform from \citet{gill-wavelet}, we found that similar quantitative results from \citet{gill-wavelet} could only be reproduced using a Mexican Hat kernel that was not normalized.  With a normalized kernel, the slope is decreased by $2$ compared to the slopes reported by \citet{gill-wavelet}.  This difference is due to the Mexican hat wavelet's close relation to the Laplacian of a Gaussian kernel.  The Gaussian kernel carries units of $\mathrm{length}^{-2}$ and its response remains constant across all scales \citep{lindeburg1994}.  Each derivative causes the response to decay by roughly $1/\sigma$, since the derivative effectively introduces an additional unit of $\mathrm{length}^{-1}$.  To correct for the decaying response, the convolution of the image and Mexican Hat wavelet should be multiplied with $\sigma^2$, restoring the $\mathrm{length}^{-2}$ units of the Gaussian kernel.  This normalization accounts for the difference of $2$ we find in the slopes. This approach is known as {\it scale-normalized derivatives} and is essential in blob detection algorithms \citep{lindeburg1994}.

\begin{figure}
\includegraphics[width=0.5\textwidth]{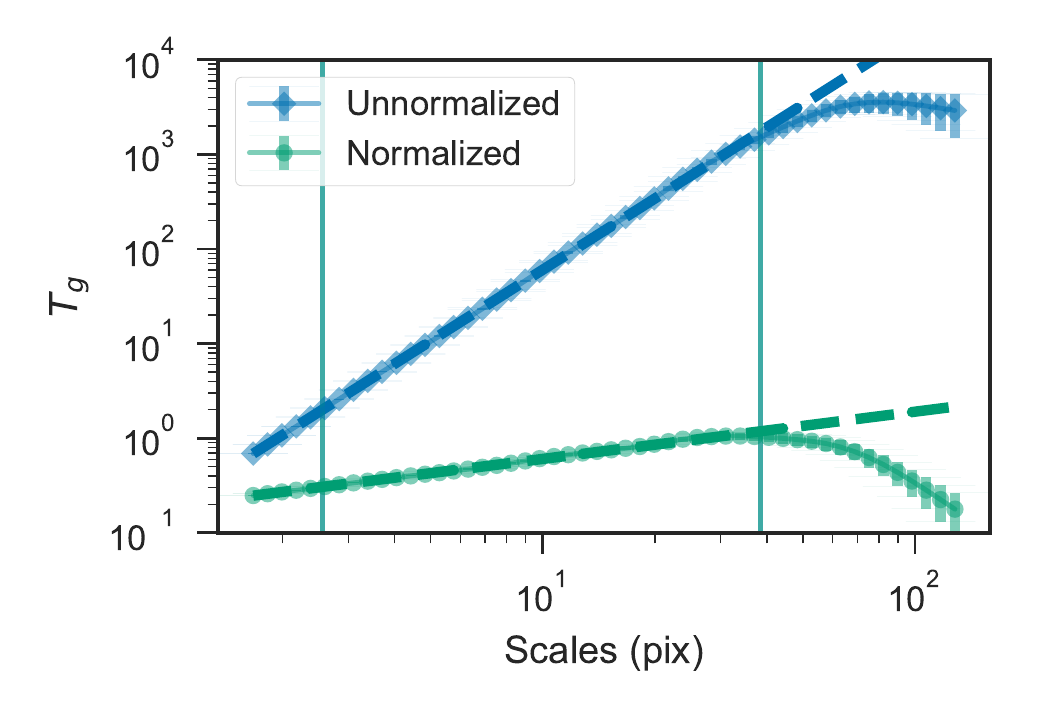}
\caption{\label{fig:wavelet_norm} Wavelet transform with (green circles) and without (blue diamonds) a normalized Mexican Hat wavelet.  The dashed lines are the fits to the respective curves, and the region used for the fit is indicated with the vertical lines.  Deviations from a single power-law relation are more significant in the normalized transform.}
\end{figure}

While the wavelet normalization does not change qualitative results and interpretation from \citet{gill-wavelet}, deviations from a power-law relation can be hidden when an unnormalized kernel is used.  Figure \ref{fig:wavelet_norm} shows the wavelet transform for the same dataset with and without normalization \citep[see also Figure 12 in][]{Boyden2016ApJ...833..233B}.  The unnormalized and normalized transforms have slopes of $2.50\pm0.01$ and $0.50\pm0.01$, respectively, when fitting on scales from 2.5 to 40 pixels.  Of note is the larger deviations in the normalized transform compared to the unnormalized transform on scales larger than 40 pixels.

Our implementation of the wavelet transform in the \turbustat\ package allows for the normalization to be disabled to so the \citet{gill-wavelet} results can be reproduced, though a warning is printed when doing so.


\section{Comparison to Delta-Variance IDL Code}
\label{app:comparison_to_deltavariance_idl_code}

\begin{figure}
\includegraphics[width=0.5\textwidth]{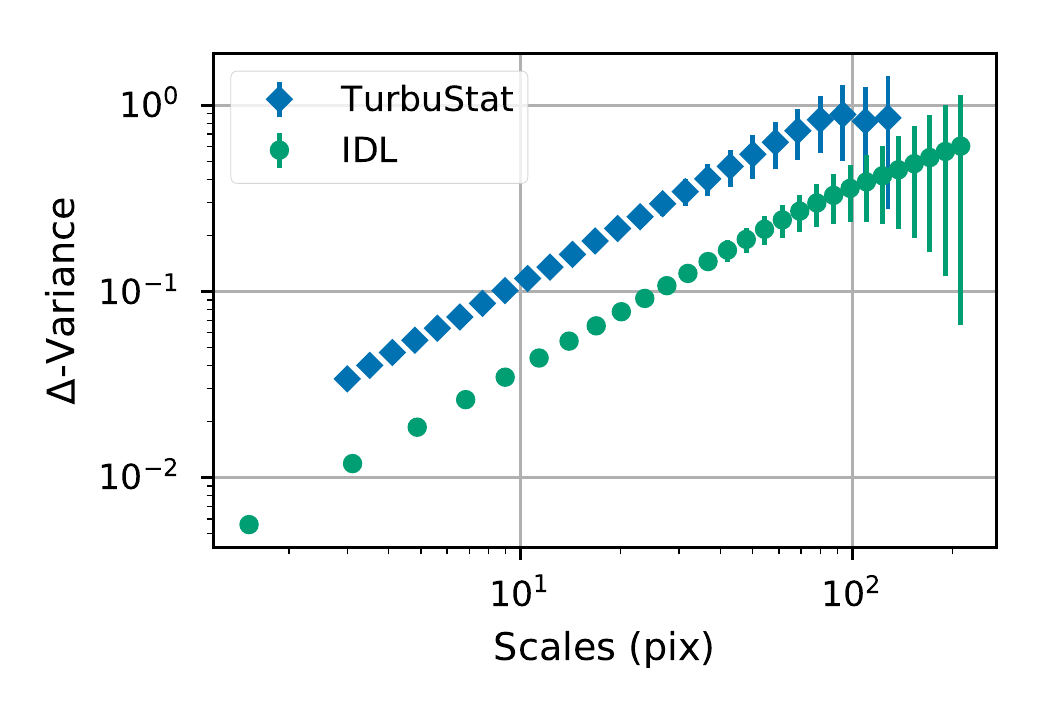}
\caption{\label{fig:delvar_idl} Delta-variance curves from \turbustat\ (blue diamonds) and the original IDL implementation described in \citet[][green circles]{oss08I-delvar} of an fBM image with an index of $-3.0$.  The recovered slope of both curves is within uncertainty of the expected slope of $1.0$.  The offset between the curves results from differences in the implementation of the convolution step.}
\end{figure}

The original version of the delta-variance described in \citet{oss08I-delvar,oss08II-delvar} is publicly available\footnote{\url{hera.ph1.uni-koeln.de/~ossk/Myself/deltavariance.html}}.
Figure \ref{fig:delvar_idl} shows the delta-variance curve from an fBM image with an index of $-3.0$, as used in \S\ref{sub:examples_with_twodim_images}.  The figures shows that the \turbustat\ implementation recovers a delta-variance curve whose slope is consistent with the IDL version, within uncertainty.  There is an offset between the curves that results from differences in the implementations of the convolution step\footnote{\turbustat\ uses astropy's FFT convolution function (\url{docs.astropy.org/en/stable/convolution}).}.  The offset is constant within uncertainty for all scales and remains the same when tested on fBM images with different indices.  Thus, the offset does not affect the recovered slope.

\section{Scaling Tests}
\label{app:scaling_tests}

We summarize a set of performance tests in this section for the 14 statistical methods in TurbuStat. These tests were run with the default settings on synthetic data ranging from sizes of $256^2$ to $2048^2$ for two-dimensional images, and $256^3$ to $2048^3$ for spectral-line data cubes.  We ran the tests on a compute node with two Intel E5-2683 v4 ``Broadwell'' processor and 512 GB\footnote{This is a ``large 512 G'' on the cedar cluster (\url{docs.computecanada.ca/wiki/Cedar}).}; the times are based on running on a single processor.

\begin{figure*}
\includegraphics[width=\textwidth]{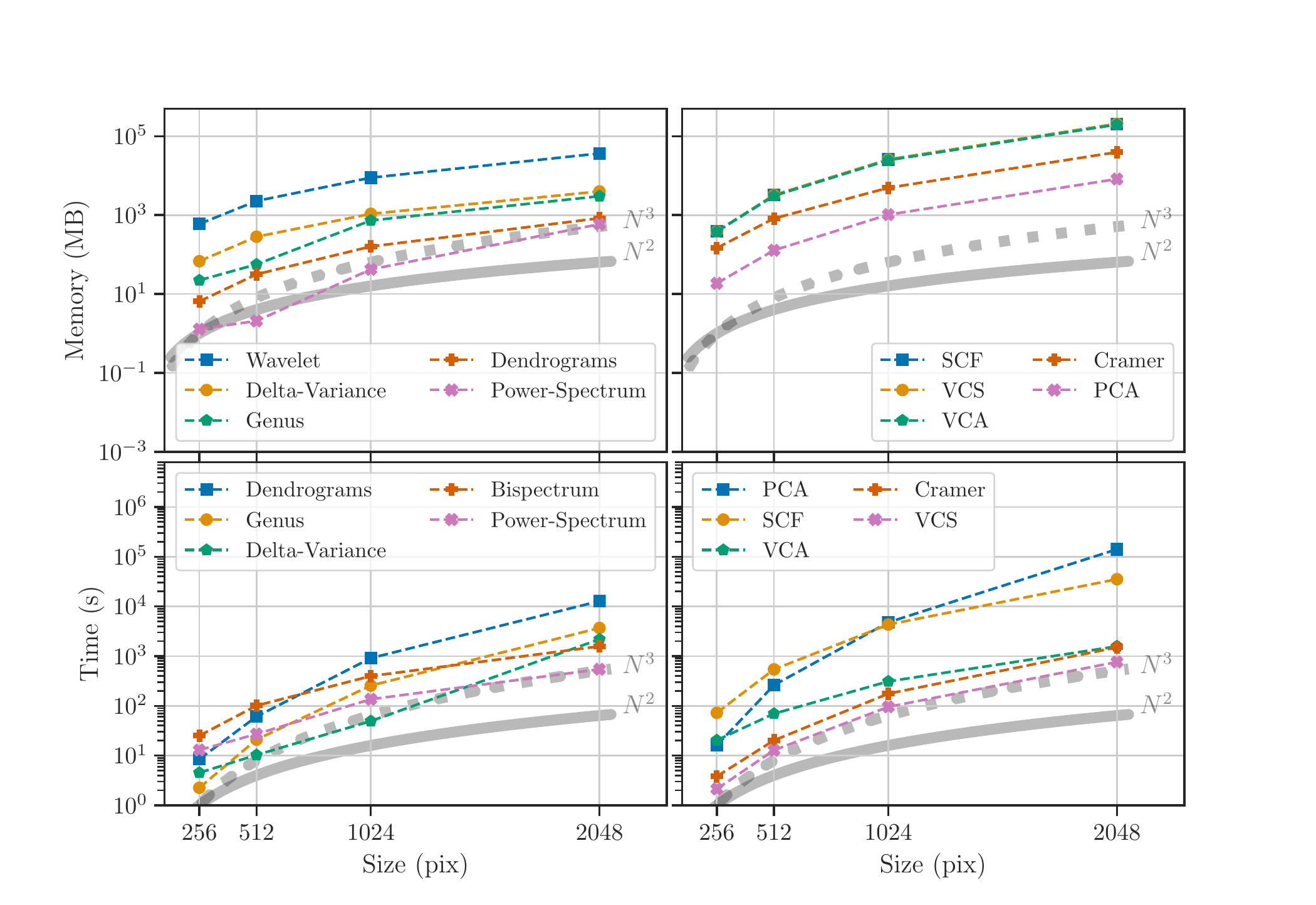}
\caption{\label{fig:scaling_test} Memory (top row) and run time (bottom row) tests for two-dimensional (left column) and three-dimensional (right column) statistical methods as a function of image/cube size in pixels. The solid and dash gray lines indicate $N^2$ and $N^3$ scaling relations for image/cube size $N$.  Each panel shows the five methods ordered from most to least usage in each panel, with the order shown in the legends. Other two-dimensional methods use less memory or have a shorter run time than those plotted.}
\end{figure*}

Figure \ref{fig:scaling_test} summarizes these tests by showing the five methods with the highest memory usage or run time.  We note that the memory tests {\it do not include the memory-usage of the data products}. Machine-readable tables are available for all methods as online material.

Scripts to reproduce these scaling tests are available at \url{https://github.com/Astroua/TurbuStat/tree/master/Examples}.

\end{document}